\documentclass[prd,twocolumn,superscriptaddress]{revtex4-1}
\usepackage[utf8]{inputenc}
\usepackage{amsmath,amsfonts,bm,color,braket,lipsum,xcolor,braket}
\usepackage[normalem]{ulem}
\usepackage{mathtools}

\renewcommand{\d}{\textrm{d}}
\newcommand{\ii}{\textrm{i}}
\renewcommand{\aa}{\textsc{a}}
\newcommand{\bb}{\textsc{b}}

\newcommand{\op}[1]{\hat{#1}}

\renewcommand{\vec}[1]{\bm{#1}}
\newcommand{\Eq}[1]{Eq.~\eqref{#1}}

\newcommand{\edu}{\color{magenta}}

\begin{document}

\title{Entanglement harvesting from multiple massless scalar fields and divergences in Unruh-DeWitt  detector models}

\author{{Allison M. Sachs}}
\email{asachs@uwaterloo.ca}
\affiliation{Institute for Quantum Computing, University of Waterloo, Waterloo, ON, N2L 3G1, Canada}
\affiliation{Dept. Applied Math., University of Waterloo, Waterloo, ON, N2L 3G1, Canada}

\author{Robert B. Mann}
\email{rbmann@uwaterloo.ca}
\affiliation{Institute for Quantum Computing, University of Waterloo, Waterloo, ON, N2L 3G1, Canada}
\affiliation{Dept. Physics and Astronomy, University of Waterloo, Waterloo, ON, N2L 3G1, Canada}
\affiliation{Perimeter Institute for Theoretical Physics, Waterloo, ON, N2L 2Y5, Canada}

\author{Eduardo Mart\'in-Mart\'inez}
\email{emartinmartinez@uwaterloo.ca}
\affiliation{Institute for Quantum Computing, University of Waterloo, Waterloo, ON, N2L 3G1, Canada}
\affiliation{Perimeter Institute for Theoretical Physics, Waterloo, ON, N2L 2Y5, Canada}
\affiliation{Dept. Applied Math., University of Waterloo, Waterloo, ON, N2L 3G1, Canada}

\begin{abstract}

\date{\today}

We analyze scenarios in which pairs of Unruh-DeWitt-like detectors interact with scalar field(s) with  bi-linear and quadratic coupling.  For all cases --  real scalar fields, charged scalar fields, and even for couplings mixing two different scalar fields -- we find that the detectors' dynamics depends on the Wightman function of a real scalar field with the same functional form up to a constant multiple.  Consequently entanglement harvesting exhibits low sensitivity to either the kind of scalar field or their charged nature. Furthermore, we show on general grounds that pairs of Unruh-DeWitt-like detectors exhibit persistent divergences when they interact with scalar field(s) with a quadratic or bi-linear coupling and discuss possible avenues to improve the models through the  regularization of such divergences.
\end{abstract}
\pacs{Valid PACS appear here}
\maketitle

\section{Introduction}

Unruh DeWitt (UDW)   detectors have been used to study the Unruh effect, Hawking radiation, the entanglement structure of various quantum field theories (QFT) as well as other quantum features of QTFs, such as vacuum fluctuations.

Primarily these studies have been confined to bosonic and real QFTs \cite{unruh:1976aa,hawking:1979aa,ng:2014aa,pozas:2015aa}. The fermionic sector has been considerably less explored, primarily due to the persistent divergences only renormalized in the last few years \cite{hinton:1984aa,Takagi:1985aa,Takagi:1986aa,hummer:2016aa}. Despite this, there are already interesting results in the fermionic sector \cite{louko:2016aa,sachs:2017aa}. 

However, studies looking into the {entanglement structure of a QFT by using a UDW detector 
(a process   called \emph{entanglement harvesting} \cite{Reznik:2003aa,Valentini:1991aa})  have revealed a surprising \emph{persistent} divergence, exhibited only in the correlations of a multi-detector system, if the detector is coupled quadratically to a real scalar field.}
A persistent divergence in a UDW model is a divergence that remains  even for a choice of detector switching function and spatial profile that are both smooth\cite{louko:2008aa,louko:2006aa,sachs:2017aa}. A full study of the entanglement harvesting protocol in a fermionic QFT will require renormalizing or otherwise removing this persistent divergence \cite{sachs:2017aa}.
 
Our purpose in this paper is  to shed light on this persistent divergence by analyzing the behaviour of different \emph{UDW-like detector models}. Models considered UDW-like are those that describe some number of non-relativistic, localized quantum systems called  \emph{detectors}, (such as qubits)  coupled to a QFT or QFTs via an interaction that has a well-defined size and shape for each detector (called the \emph{spatial smearing} and \emph{switching function}  respectively).  Both single detectors \cite{unruh:1976aa,hawking:1979aa,louko:2016aa} and pairs of detectors \cite{Valentini:1991aa,Reznik:2003aa,Reznik:2005aa,pozas:2016aa} are commonly studied in the literature. It is less common that more detectors are considered \cite{Kukita:2017aa}.

It is in the case of pairs of detectors that one can study the phenomenon of entanglement harvesting \cite{pozas:2015aa}, entanglement farming \cite{martinmartinez:2013aa}, and harvesting-based ways of distinguishing different background geometries \cite{VerSteeg2009,salton:2015aa,ng:2017aa}  and topologies \cite{martinmartinez:2016aa}. Furthermore, in certain setups, pairs of detectors exhibit persistent divergences, while individual detectors were found to be free of such divergences \cite{sachs:2017aa}.

Here we investigate the behaviour of a pair of UDW-like detectors, each of which couple multilinearly to 
a collection of scalar quantum fields. Our purpose is to study the structure of persistent divergences in
non-local correlations that
were first discovered \cite{sachs:2017aa} for detectors coupling  quadratically to a single field operator of a scalar QFT  (we refer to these as  \emph{quadratic} UDW detectors). We find that such persistent divergences generically appear for bilinear couplings, and that these have the same form as those exhibited in the quadratic case.    We expect these divergence to appear for any  multilinear coupling to scalars.

We begin   in section \ref{sec:setup} by setting up the formalism for multilinearly coupled UDW-like detectors to a collection of scalar fields. In Section~\ref{sec:pairs} we will particularize to pairs of detectors, from which the single detector model is easily recovered. The analysis in section Section~\ref{sec:results} will focus on models that are closely related to the divergent model discussed in \cite{sachs:2017aa}. Indeed, the discussion in Section~\ref{sec:discussion} will focus primarily on understanding when these divergences will be present in these related models.

\section{Setup\label{sec:setup}}

  We consider a \textit{detector} to be a first-quantized quantum system that couples locally to a quantum field. Inspired by particle detector models such as that by Unruh-DeWitt \cite{hawking:1979aa}, and non-linear variants of this model \cite{Takagi:1986aa,Takagi:1985aa,hinton:1984aa,hummer:2016aa}, let us consider  the following interaction picture Hamiltonian coupling detector observables $\op D_\gamma(\vec{x},t)$ to scalar fields in the following very general form:  
\begin{align}
    \op H(t)=\sum_{\gamma\in\mathbb{D}}\int \d^d\vec{x}\op D_\gamma(\vec{x},t)
	\, \prod_{i=1}^{n}
    f^\gamma_i(\op \phi_i(\vec x, t)),
\label{multifieldham}
\end{align}
where $\mathbb{D}=\{A,B\}$ and $\op \phi_i(\vec x, t)$ are the field operators for 
scalar fields (indexed by $i=1,\ldots,n$);  $f^\gamma_i$ is a (not necessarily linear) function of $\op \phi_i(\vec x, t)$.

For our analysis   we will choose the operators $\op D_\gamma$ to correspond to a set of monopole densities of comoving detectors (sharing a centre-of-mass proper time $t$), as in the UDW model. This  can be written as a product in the following way 
\begin{align}
    \op D_\gamma(\vec{x},t):=\lambda_\gamma\op \mu_\gamma (t)\chi_\gamma (t)F_\gamma (\vec x),
\end{align}
where $\chi_\gamma (t)$ and $F_\gamma (\vec x)$ describe the switching and spatial shape (in their co-moving reference frames) of the first-quantized  detectors labeled by $\gamma$. The operator $\op \mu_\gamma(t)$   accounts for the internal structure of the monopole moment of the $\gamma-$th detector, which we reduce to two levels as is common in UDW models, i.e.
\begin{align}
    \op \mu_\gamma (t):=e^{-\ii\Omega_\gamma t}\ket{g_\gamma}\bra{e_\gamma}+e^{\ii\Omega_\gamma t}\ket{e_\gamma}\bra{g_\gamma},
\end{align}
where where $\ket{g_\gamma}$ represents the free ground eigenstate of the detector $\gamma$ and $\ket{e_\gamma}$ represents the excited state, separated from the ground state by a possibly-vanishing energy gap $\Omega$.

We consider first the case of an arbitrary number of detectors indexed by the set $\mathbb{D}$. This motivation for studying this particular multi-field Hamiltonian comes from the divergence found in the entanglement of a pair of Unruh Dewitt detectors coupled quadratically to a massless scalar field, as in \cite{sachs:2017aa}. To attain that particular Hamiltonian from \Eq{multifieldham}, we set  $\chi$ and $F$ to Gaussian distributions. \ Let there be a single scalar field $\op \phi_1(\vec x, t)$, and set the function $f^\gamma_1(\op\phi_1)=:\op \phi_1^2(\vec x, t):$. Explicitly, the interaction Hamiltonian   is \cite{sachs:2017aa}  
\begin{align}
    \op H(t)=\sum_{\gamma\in\{A,B\}}\int \d^d\vec{x}\op \mu_\gamma (t)\chi_\gamma (t)F_\gamma (\vec x):\op \phi^2(\vec x, t): 
\label{quadfieldham1}
\end{align}
letting $\op \phi_1(\vec x, t)=\op \phi(\vec x, t)$.

We will ask if/when the divergence studied in \cite{sachs:2017aa} shows up with a Hamiltonian of the form \Eq{multifieldham}, and, if so, does it have the same structure as in the quadratic single-field case \Eq{quadfieldham1}.

Derivations for finding the time-evolved state of a two UDW-detector system coupled to various single-field couplings are straightforward and can be found throughout the literature (see e.g., \cite{pozas:2015aa, sachs:2017aa}). While generalization to many-detector systems is a simple extension \cite{Kukita:2017aa},  the extension to multiple fields is novel. The notation used most closely follows \cite{pozas:2015aa}, excepting the sum/product notation to condense the expressions with many Hilbert spaces, which was not necessary in previous investigations.

\subsection{State of system after interaction}

In this section we will derive a very general expression for the state of the combined field-detectors system. We begin by following a similar process as in \cite{pozas:2015aa,sachs:2017aa}.  Starting with the interaction picture time evolution operator $\op U=\mathcal{T}\exp(-\ii{\int \d t \op H_I(t) })$, and taking a Dyson expansion we obtain
\begin{align} 
	\op U = \openone +\op U^{(1)}+\op U^{(2)}+\mathcal{O}(\lambda^3_\gamma) \label{expansion}
\end{align}
where
\begin{align} 
	\op U^{(1)} &= -\ii  \! 
	\int_{-\infty}^\infty \!\!\!\! \d t  \, 
        \op H_\text{I}(t)     \label{u1}      \\
        \op U^{(2)} &=  - \! 
    \int_{-\infty}^{\infty} \!\!\!\!  \d t    
    \int_{-\infty}^{t} \!\!\!\!\!   \d t'  \, 
        \op H_\text{I}(t)
        \op H_\text{I}(t') \label{u2}.
\end{align}
Using  the Hamiltonian \eqref{multifieldham}, these operators can be equivalently written as
\begin{align} 
	\op U^{(1)} &= 
    	-\ii  \! 
    	\sum_{\gamma\in\mathbb{D}}
    	\int_{-\infty}^\infty \!\!\!\! \d t  \!
            \int \d^d\vec{x}
            \op D_\gamma(\vec{x},t)
            \prod_{i=1}^{n}
            f^\gamma_i(\op \phi_i(\vec x, t))   \\
	\op U^{(2)} =  &
        - \! 
        \sum_{\gamma,\nu\in\mathbb{D}}
        \int_{-\infty}^{\infty} \!\!\!\!\!\!  \d t \!\!
        \int_{-\infty}^{t} \!\!\!\!\!\!\!   \d t'  \!\! 
        \int \!\!\d^d\vec{x} \!\!
        \int \!\!\d^d\vec{x}'
            \op D_\gamma(\vec{x},t)
            \op D_\nu(\vec{x}',t')
        \notag\\&\quad\times
            \prod_{i,j=1}^{n}
            f^\gamma_i(\op \phi_i(\vec x, t))
            f^\nu_j(\op \phi_j(\vec x', t')) 
\end{align}
and $\hat U^{(0)}=\openone$.

Given an initial state $\op \rho_0$, we can express its time evolution under  $\op U$ in a perturbative expansion. We will call this time evolved state simply $\op \rho$; using \eqref{expansion} we obtain
\begin{align} 
	&\op \rho=\op \rho_0 \! + \! \op \rho^{(1,0)}  \! + \! \op \rho^{(0,1)}  \! + \! \op \rho^{(1,1)}  \! + \! \op \rho^{(2,0)}  \! + \! \op \rho^{(0,2)}  \! +\mathcal{O}(\lambda^3_\gamma)
\label{rho1}
\end{align}
where $\op \rho^{(i,j)} =\op U^{(i)} \op \rho_{0} \op U^{(j)\dagger}$.  

Taking the ground state to be the initial state of each detector system and choosing arbitrary pure, uncorrelated states for the field systems 
\begin{align}
    \op\rho_0=\bigotimes_{\delta\in\mathbb{D}}
        \ket{g}_\delta\prescript{}{\delta}{\bra{g}}
    \bigotimes_{i=1}^{n}
        \ket{\Psi_i}\bra{\Psi_i},
\end{align}
we find  that \Eq{rho1} can be written as
\begin{align}
    \op \rho^{\phi}=&\op\rho_0^{\phi}+
        \sum_{\gamma}
            \Big(
                \lambda_{\gamma}\op\rho_{\gamma}^{\phi}
                +\lambda^2_{\gamma}\,\op\rho_{\gamma\gamma}^{\phi}
            \Big)
            \bigotimes_{\substack{\delta\in\mathbb{D}\\\delta\ne\gamma}}
            \ket{g}_\delta\prescript{}{\delta}{\bra{g}}
        \notag\\&
        +\!\sum_{\substack{\gamma,\nu\in\mathbb{D}\\\gamma\ne\nu}}\!
            \Big(
                \lambda_{\gamma}\lambda_{\nu}\,\op\rho_{\gamma\nu}^{\phi}
            \Big)
            \!\bigotimes_{\substack{\delta\in\mathbb{D}\\\delta\ne\gamma,\nu}}\!
            \ket{g}_\delta\prescript{}{\delta}{\bra{g}}
        +\mathcal{O}(\lambda_{\gamma\nu}^3),\label{readibleWithFields}
\end{align}
where  
\begin{align}
    \lambda_\gamma\op\rho_{\gamma}^{\phi}&=-\ii
    	\int_{-\infty}^\infty \!\!\!\! \d t  \!
        \int \d^d\vec{x}
            \op D_\gamma(\vec{x},t)
            \ket{g}_\gamma\prescript{}{\gamma}{\bra{g}}
    \notag\\&\quad\times
            \bigotimes_{i=1}^{n}
            f^\gamma_i(\op \phi_i(\vec x, t))
            \ket{\Psi_i}\bra{\Psi_i}
            \,+\,\text{H.c.},\label{1stOrderRespose}
\end{align}
and $\rho_{\gamma\nu}^\phi$ is defined the following way if $\gamma\ne\nu$
\begin{align}
    \lambda_\gamma&\lambda_\nu\op\rho_{\gamma\nu}^{\phi}= 
    \Big(
    	\!\sum_{\substack{\gamma,\nu\in\mathbb{D}\\\gamma\ne\nu}}\!
    	\int_{-\infty}^\infty \!\!\!\! \d t  \!
    	\int_{-\infty}^\infty \!\!\!\! \d t'  \!
        \int \d^d\vec{x}
        \int \d^d\vec{x}'
    \notag\\&
    \quad\times
        \op D_\gamma(\vec{x},t)
        \ket{g}_\gamma\prescript{}{\gamma}{\bra{g}}
        \otimes
        \ket{g}_\nu\prescript{}{\nu}{\bra{g}}
        \op D_\nu(\vec{x}',t')
    \notag\\&
    \quad\times
            \bigotimes_{i=1}^{n}
            f^\gamma_i(\op \phi_i(\vec x, t))
            \ket{\Psi_i}\bra{\Psi_i}
            f^\gamma_i(\op \phi_i(\vec x', t'))\Big)
    \notag\\&
    -\Big(
        \sum_{\substack{\gamma,\nu\in\mathbb{D}\\\gamma\ne\nu}}
        \int_{-\infty}^{\infty} \!\!\!\!\!\!  \d t \!\!
        \int_{-\infty}^{t} \!\!\!\!\!\!\!   \d t'  \!\! 
        \int \!\!\d^d\vec{x} \!\!
        \int \!\!\d^d\vec{x}'
    \label{2ndOrderCor}\\&
    \quad\times
        \op D_\gamma(\vec{x},t)
        \ket{g}_\gamma\prescript{}{\gamma}{\bra{g}}
        \otimes
        \op D_\nu(\vec{x}',t')
        \ket{g}_\nu\prescript{}{\nu}{\bra{g}}
    \notag\\&
    \quad\times
        \bigotimes_{i=1}^{n}
        f^\gamma_i(\op \phi_i(\vec x, t))
        f^\gamma_i(\op \phi_i(\vec x', t'))
        \ket{\Psi_i}\bra{\Psi_i}
        +\text{H.c.}\Big), \notag
\end{align}
and as the following when $\gamma=\nu$
\begin{align}
    \lambda_\gamma^2&\op\rho_{\gamma\gamma}^{\phi}=
    \Big(
    	\sum_{\gamma\in\mathbb{D}}
    	\int_{-\infty}^\infty \!\!\!\! \d t  \!
    	\int_{-\infty}^\infty \!\!\!\! \d t'  \!
        \int \d^d\vec{x}
        \int \d^d\vec{x}'
    \notag\\&
    \quad\times
        \op D_\gamma(\vec{x},t)
        \ket{g}_\gamma\prescript{}{\gamma}{\bra{g}}
        \op D_\gamma(\vec{x}',t')
    \notag\\&
    \quad\times
            \bigotimes_{i=1}^{n}
            f^\gamma_i(\op \phi_i(\vec x, t))
            \ket{\Psi_i}\bra{\Psi_i}
            f^\gamma_i(\op \phi_i(\vec x', t'))\Big)
    \notag\\&
    -\Big(
        \sum_{\gamma\in\mathbb{D}}
        \int_{-\infty}^{\infty} \!\!\!\!\!\!  \d t \!\!
        \int_{-\infty}^{t} \!\!\!\!\!\!\!   \d t'  \!\! 
        \int \!\!\d^d\vec{x} \!\!
        \int \!\!\d^d\vec{x}'
    \label{2ndOrderRespose}\\&
    \quad\times
        \op D_\gamma(\vec{x},t)
        \op D_\gamma(\vec{x}',t')
        \ket{g}_\gamma\prescript{}{\gamma}{\bra{g}}
    \notag\\&
    \quad\times
        \bigotimes_{i=1}^{n}
        f^\gamma_i(\op \phi_i(\vec x, t))
        f^\gamma_i(\op \phi_i(\vec x', t'))
        \ket{\Psi_i}\bra{\Psi_i}
        +\text{H.c.}\Big). \notag
\end{align}

The form \Eq{readibleWithFields} of the time evolved density matrix is particularly intuitive, since it separates  terms that are local to single detectors  (\Eq{1stOrderRespose} and \Eq{2ndOrderRespose}), from terms containing  correlations  that the detectors acquire through their interaction with the field, which can be seen in \Eq{2ndOrderCor}.

\subsection{State of detectors after interaction}

We now turn to the problem of finding a general expression for the combined state of the detectors by tracing out all field degrees of freedom.   Tracing over the field in \Eq{readibleWithFields}, we obtain  
\begin{align}
    \op \rho=&\op\rho_0+
        \sum_{\gamma}
            \Big(
                \lambda_{\gamma}\op\rho_{\gamma}
                +\lambda^2_{\gamma}\,\op\rho_{\gamma\gamma}
            \Big)
            \bigotimes_{\substack{\delta\in\mathbb{D}\\\delta\ne\gamma}}
            \ket{g}_\delta\prescript{}{\delta}{\bra{g}}
        \notag\\&
        +\!\sum_{\substack{\gamma,\nu\in\mathbb{D}\\\gamma\ne\nu}}\!
            \Big(
                \lambda_{\gamma}\lambda_{\nu}\,\op\rho_{\gamma\nu}
            \Big)
            \!\bigotimes_{\substack{\delta\in\mathbb{D}\\\delta\ne\gamma,\nu}}\!
            \ket{g}_\delta\prescript{}{\delta}{\bra{g}}
        +\mathcal{O}(\lambda_{\gamma\nu}^3)\label{readible},
\end{align}
where dropping the superscript $\phi$ indicates  that the trace over the field has been performed. The operator $\op\rho_{\gamma}$ is defined by

\begin{align}
    \op\rho_{\gamma}&=-\ii
    	\int_{-\infty}^\infty \!\!\!\! \! \d t  \!
        \int \d^d\vec{x}
            L_{\gamma} (t,\bm{x})
            \prod_{i=1}^{n}
            S_{\Psi_i}^{f^\gamma_i}(\vec x, t)
            \ket{e}_\gamma\prescript{}{\gamma}{\bra{g}}
            \,+\,\text{H.c.}
\end{align}
where $L_{\nu}$ 
is given by
\begin{align} 
	L_{\nu} (t,\bm{x})=& \,
		\chi_{\nu}(t-t_\nu) \, F_{\nu}(\bm{x}-\bm{x}_\nu)  \, e^{\ii \Omega_{\nu} t}\label{linearlittleL}
\end{align}
and $S$ is the single point function of $f^\gamma_i(\phi_i(\vec x, t))$
\begin{align}
    S_{\Psi_i}^{f^\gamma_i}(\vec x, t)=\bra{\Psi_i}f^\gamma_i(\phi_i(\vec x, t))\ket{\Psi_i}\label{onepoint},
\end{align}
and $\rho_{\gamma\nu}$ is defined as       
\begin{align}\label{chdata}
    \op\rho_{\gamma\nu}&=
    	\!\sum_{\substack{\gamma,\nu\in\mathbb{D}\\\gamma\ne\nu}}\!
    	\int_{-\infty}^\infty \!\!\!\!\! \d t  \!\!
    	\int_{-\infty}^\infty \!\!\!\!\! \d t' \!\!
        \int \!\!\d^d\vec{x}\!\!
        \int \!\!\d^d\vec{x}'
        \left(\mathcal{L} _{\gamma\nu}^{f^\gamma_i}(\vec x, t,\vec x', t')\right)^*
    \notag\\&
    \quad\times
 {\ket{e_\gamma g_\nu}\bra{g_\gamma e_\nu}}   
    \notag\\&
    -\Big(
        \Big(\sum_{\substack{\gamma,\nu\in\mathbb{D}\\\gamma\ne\nu}}
        \int_{-\infty}^{\infty} \!\!\!\!\!\!  \d t \!\!
        \int_{-\infty}^{t} \!\!\!\!\!\!\!   \d t'  \!\! 
        \int \!\!\d^d\vec{x} \!\!
        \int \!\!\d^d\vec{x}'
        \mathcal{M}_{\gamma\nu}^{f^\gamma_i}(\vec x, t,\vec x', t')
    \notag\\&
    \quad\times
      {\ket{e_\gamma e_\nu}\bra{g_\gamma g_\nu}}  \Big)
        +\text{H.c.}\Big) 
\end{align}
if $\gamma\ne\nu$, and as 
\begin{align}
    \op\rho_{\gamma\gamma}&=
    	\sum_{\gamma\in\mathbb{D}}
    	\int_{-\infty}^\infty \!\!\!\!\! \d t  \!\!
    	\int_{-\infty}^\infty \!\!\!\!\! \d t' \!\!
        \int \!\!\d^d\vec{x}\!\!
        \int \!\!\d^d\vec{x}'
        \left(\mathcal{L} _{\gamma\gamma}^{f^\gamma_i}(\vec x, t,\vec x', t')\right)^*
    \notag\\&
    \quad\times
        \left( {\ket{e_\gamma g_\gamma}\bra{g_\gamma e_\gamma}}  \right)
\end{align}
when $\gamma=\nu$.

In order to align our notation with that in the literature  \cite{pozas:2016aa}, we have defined the following functions  $\mathcal{M}_{\gamma\nu}$ and $\mathcal{L}_{\gamma\nu}$:
\begin{align} 
	\mathcal{M}_{\gamma\nu}^{f^\gamma_i} \! =	&-\lambda_{\aa}\lambda_{\bb}
		\int_{-\infty}^{\infty} \!\!\!\!\! \d t  
		\int_{-\infty}^{t} \!\!\!\!\! \d t' \!
		\int \d^d \bm{x}  
		\int  \d^d \bm{x}' \, 
		\notag\\ & \times\label{generalM}
		L_{\gamma} (t,\bm{x})
        L_{\nu} (t'\!,\bm{x}') \,
		\prod_{i=1}^{n}
		W^{f^\gamma_i}_{\ket{\Psi_i}} (t,\bm{x},t'\!,\bm{x}')\\[3mm]
	\mathcal{L} _{\gamma\nu}^{f^\gamma_i} \! = & \lambda_\nu\lambda_\mu
		\int_{-\infty}^{\infty} \!\!\!\!\! \d t  
		\int_{-\infty}^{\infty} \!\!\!\!\! \d t' \!
		\int  \d^d \bm{x}  
		\int  \d^d \bm{x}' \,
		\label{generalL}\\ & \times\notag
		L^*_{\gamma} (t ,\bm{x} ) \,
		 L_{\nu} (t'\!,\bm{x}') \, 
		\prod_{i=1}^{n}
		W^{f^\gamma_i}_{\ket{\Psi_i}} (t,\bm{x},t'\!,\bm{x}'),
\end{align}
where L is defined in \Eq{linearlittleL} and we have utilized the cyclic property to re-express the trace of the field as either the single point function of $f^\gamma_i(\phi_i(\vec x, t))$ (defined in \Eq{onepoint}) or the two-point correlator
\begin{align}
    W_{\Psi_i}^{f^\gamma_i}(\vec x, t,\vec x', t')=\bra{\Psi_i}f^\gamma_i(\phi_i(\vec x, t))f^\gamma_i(\phi_i(\vec x', t'))\ket{\Psi_i}.
\end{align}

The term $\mathcal{M}$ is associated with non-local correlations, while $\mathcal{L}_{\gamma\nu}$ is associated with mutual information   including classical correlations. and $\mathcal{L}_{\gamma\gamma}$   is each detector's local excitation probability that we can associate with local `noise' in the context of entanglement harvesting \cite{pozas:2015aa}.

\section{Analysis of pairs of detectors\label{sec:pairs}}
In this section we particularize to pairs of detectors. Apart from being the minimal setup for performing entanglement harvesting, this will provide us with better insight into understanding the divergence behaviour of different UDW-like detector models. In particular, in Section~\ref{sec:discussion} we will be able to focus on whether divergences will appear in models similar to the quadratically-coupled model, which exhibits persistent divergences when pairs of detectors are considered \cite{sachs:2017aa}.

We begin our study of pairs of detectors first by noting the equivalence between two scenarios. The first is a setup with more than two detectors, but in which all but two detectors have been traced out. The second  scenario is a setup that initially only has two detectors. For the initial state chosen for the detectors, these scenarios are, at leading order, equivalent since the multi-partite contributions to the detector state are all traceless when all detectors begin in energy eigenstates. Thus,  after tracing out detector $\gamma$, any term involving $\gamma$ will also vanish. In other words,  at second order, the field(s) only mediate bipartite detector interactions. In this section we will reduce our analysis to a pair of detectors, A and B, regardless of whether they are contained in some possibly-larger set of detectors.

We find  
\begin{align}
    \op \rho_{\aa\bb}=&\op\rho_{0,\aa\bb}+
        \lambda_{\aa}\op\rho_{\aa}
        +\lambda^2_{\aa}\,\op\rho_{\aa\aa}
        +\lambda_{\bb}\op\rho_{\bb}
        +\lambda^2_{\bb}\,\op\rho_{\bb\bb}
    \notag\\&
        +\lambda_{\aa}\lambda_{\bb}\,\op\rho_{\aa\bb}
        +\lambda_{\bb}\lambda_{\aa}\,\op\rho_{\bb\aa}
    +\mathcal{O}(\lambda_{\gamma\nu}^3)\label{readibleABtwo},
\end{align}
for the time-evolved state of pairs of detectors, where
\begin{align}
    \op\rho_{0,\aa\bb}=\mathrm{Tr}_{\gamma\ne \aa,\bb}(\op\rho_0)=\ket{g}_{\aa}\prescript{}{\aa}{\bra{g}}
    \bigotimes\ket{g}_{\bb}\prescript{}{\bb}{\bra{g}}.
\end{align}

If the initial state of the fields has a vanishing one-point function (for example, if we start in the vacuum, or any convex combination of Fock states) the first order contribution vanishes, and we can represent the detector-pair density matrix as
\begin{align} 
	\op \rho^{f^\gamma_i}_{{\aa\bb}} = \! 
		\begin{pmatrix}
			 1\!-\!\mathcal{L}^{f^\gamma_i}_{\aa\aa}\!\!-\!\mathcal{L}^{f^\gamma_i}_{\bb\bb}\!\!\! & 0 & 0 & \left(\mathcal{M}^{f^\gamma_i} \right)^* \\
			 0 & \!\mathcal{L}^{f^\gamma_i}_{\aa\aa} &  \mathcal{L}^{f^\gamma_i}_{\aa\bb}\! & 0 \\
			 0 & \!\mathcal{L}^{f^\gamma_i}_{\bb\aa} &  \mathcal{L}^{f^\gamma_i}_{\bb\bb}\!  & 0 \\
			 \mathcal{M}^{f^\gamma_i}  & 0 & 0 & 0 \\
		\end{pmatrix} \! \!
		+\!\mathcal{O}(\lambda^4_\mu)\label{densitymatrix},
\end{align}
where $\mathcal{M}^{f^\gamma_i}=\mathcal{M}_{\aa\bb}^{f^\gamma_i}+\mathcal{M}_{\bb\aa}^{f^\gamma_i}$ (to match the notation in the literature), where $\mathcal{M}_{\gamma\nu}^{f^\gamma_i}$ and $\mathcal{L}_{\gamma\nu}^{f^\gamma_i}$ are those in \Eq{generalM} and \Eq{generalL}, respectively, and where the following basis elements are used:
\begin{align}\label{notationbasis}
    \ket{g_\aa g_\bb}&=\begin{pmatrix}1\\0\\0\\0\end{pmatrix},\quad
    \ket{g_\aa e_\bb}=\begin{pmatrix}0\\1\\0\\0\end{pmatrix},\notag\\
    \ket{e_\aa g_\bb}&=\begin{pmatrix}0\\0\\1\\0\end{pmatrix},\quad
    \ket{e_\aa e_\bb}=\begin{pmatrix}0\\0\\0\\1\end{pmatrix}.
\end{align}

To see the form of this matrix in general when the one-point function of (a single) field does not vanish one can check, e.g., equation (55) of \cite{simidzija:2017aa}.

What we can see is that pairs of detectors, including when they are a subset of the total collection of detectors in consideration, reproduce the usual ``X-state" found in the literature \cite{martinmartinez:2016aa,pozas:2016aa}.

\section{Results\label{sec:results}}

In this section, we will consider four different cases of \Eq{multifieldham}, two of which are known in the literature. The two known cases are the usual Unruh-DeWitt model discussed in Section~\ref{sec:usual} \cite{pozas:2015aa} and the Unruh-DeWitt-like detector coupled quadratically to a real field discussed in Section~\ref{sec:quad} \cite{hummer:2016aa,sachs:2017aa}. We will subsequently calculate the same quantities for two additional setups of Unruh-DeWitt-like detectors: in Section~\ref{sec:complex} for detectors coupled quadratically to a complex field, and in Section~\ref{sec:bilinear} for  detectors   coupled to two real fields in a bilinear manner.

  Let us consider all fields to be initialized to their respective vacuum states, represented by the vector $\ket{\Omega_i}$. This causes the single point function to vanish \cite{pozas:2015aa}, such that $\op\rho_\gamma$ vanishes for all $\gamma\in\mathbb{D}$. Thus, in this section, the only quantities of interest are the two-point correlators of $f^\gamma_i(\op\phi_i(\vec x, t))$ (or $f^\gamma_i(\op\Phi_i(\vec x, t))$ in the complex case). In the next four sections, we will relate these two-point correlators to the Wightman function of the real scalar field. This will allow, in Section~\ref{sec:discussion}, a comparison of the four models, with a particular focus on comparing the complex and bilinear models to the existing real, quadratic model   studied in \cite{hummer:2016aa}.

\subsection{Linear coupling to single real scalar field (the ususal unrh-dewit model)\label{sec:usual}}

The linear coupling considered in this section is the standard Unruh-DeWitt model found in the literature \cite{pozas:2015aa}.   This model captures the fundamental features of the light-matter interaction when the exchange of angular momentum does not play a relevant role \cite{pozas:2016aa,martinmartinez:2018aa}, and  has been used extensively for studying the Unruh effect, Hawking radiation \cite{unruh:1976aa}, 
and entanglement harvesting \cite{salton:2015aa,pozas:2015aa}, and, relevant to our later discussion regarding divergent models, the usual Unruh-DeWitt model has been shown to be free of persistent divergences \cite{satz:2007aa,louko:2006aa,louko:2008aa}. 

The Hamiltonian for a pair of Unruh-DeWitt detectors coupled linearly to a real scalar field is
\begin{align}
    \op H(t)=\sum_{\gamma\in\{A,B\}}\int \d^d\vec{x}\op D_\gamma(\vec{x},t)
    \op \phi(\vec x, t),
\label{UDWfieldham}
\end{align}
which can be attained from \Eq{multifieldham} by setting $n=1$ and letting 
\begin{align}
    f^\gamma_1(\op\phi_1(\vec x,t))=\op\phi_1(\vec x,t).
\end{align}
In this case, the two-detector density matrix is characterized by  
the vacuum Wightman function 
\begin{align}
    W_{\Omega_1}^{ \phi}(\vec x, t,\vec x', t')=\bra{\Omega_1}\phi_1(\vec x, t)\phi_1(\vec x', t'))\ket{\Omega_1}.
\end{align}

More explicitly, the two-detector density matrix after this interaction depends on only two expressions
\begin{align} 
	\mathcal{M}^{ \phi} \! =	&-\lambda_{\aa}\lambda_{\bb}
		\int_{-\infty}^{\infty} \!\!\!\!\! \d t  
		\int_{-\infty}^{t} \!\!\!\!\! \d t' \!
		\int \d^d \bm{x}  
		\int  \d^d \bm{x}' \, 
		\notag\\ & \times\label{linM}
		L_{\gamma} (t,\bm{x})
        L_{\nu} (t'\!,\bm{x}') \,
		W_{\Omega_1}^{ \phi}(\vec x, t,\vec x', t')\\[3mm]
	\mathcal{L}^{ \phi}_{\gamma\nu} \! = & \lambda_\nu\lambda_\mu
		\int_{-\infty}^{\infty} \!\!\!\!\! \d t  
		\int_{-\infty}^{\infty} \!\!\!\!\! \d t' \!
		\int  \d^d \bm{x}  
		\int  \d^d \bm{x}' \,
		\label{linL}\\ & \times\notag
		L^*_{\gamma} (t ,\bm{x} ) \,
		L_{\nu} (t'\!,\bm{x}') \, 
		W_{\Omega_1}^{ \phi}(\vec x, t,\vec x', t').
\end{align}
Previous analyses of these expressions show that they are free of persistent divergences \cite{hummer:2016aa,sachs:2017aa}.

\subsection{Quadratic coupling to single real scalar field\label{sec:quad}}

A UDW-like detector coupling quadratically to a real scalar field has been studied previously \cite{hummer:2016aa} in comparison to the usual UDW model and to a UDW-like detector coupled to a fermionic field, and has been shown to exhibit persistent divergences when pairs of detectors are considered \cite{sachs:2017aa}.  Here we present the divergent term $\mathcal{M}$ along with finite terms $\mathcal{L}$. This is done with the goal of drawing conclusions about models whose divergence properties have not yet been studied. 

In particular we will compare the quadratic model to both the complex model and the so-called `bilinear' model. The interaction part of the Hamiltonian is
\begin{align}
    \op H(t)=\sum_{\gamma\in\{A,B\}}\int \d^d\vec{x}\op D_\gamma(\vec{x},t)
    :\op \phi^2(\vec x, t):,
\label{quadfieldham2}
\end{align}
so that 
\begin{align} 
    f^\gamma_1(\op\phi_1(\vec x,t))=:\op\phi_1^2(\vec x,t):, 
    \label{quadfunct}
\end{align}
in  \Eq{multifieldham}, thereby reproducing the quadratic coupling studied previously \cite{sachs:2017aa};  the colon notation indicates normal ordering as defined in Appendix \ref{ap:complex}.

It has previously been shown \cite{sachs:2017aa}  that the two-point correlator of \eqref{quadfunct}  can be re-expressed as 
\begin{align}
    W_{\Omega_1}^{:\phi^2:}(\vec x, t,\vec x', t')=2 \left(W_{\Omega_1}^{ \phi}(\vec x, t,\vec x', t')\right)^2,
\end{align}
or in other words in terms of the two-point correlator of $\op\phi_1(\vec x,t)$, 
where $\ket{\Omega_i}$ indicates the vacuum state of the field indexed by i. 

As is the case with the usual Unruh-DeWitt detector, the two-detector density matrix after this interaction is entirely characterized by two expressions
\begin{align} 
	\mathcal{M}^{:\phi^2:}_{\gamma\nu} \! =	&-2\lambda_{\aa}\lambda_{\bb}
		\int_{-\infty}^{\infty} \!\!\!\!\! \d t  
		\int_{-\infty}^{t} \!\!\!\!\! \d t' \!
		\int \d^d \bm{x}  
		\int  \d^d \bm{x}' \, 
		\notag\\ & \times\label{quadM}
		L_{\gamma} (t,\bm{x})
        L_{\nu} (t'\!,\bm{x}') \,
		\left(W_{{\Omega_1}}^{:\phi^2:} (t,\bm{x},t'\!,\bm{x}')\right)^2\\[3mm]
	\mathcal{L}^{:\phi^2:}_{\gamma\nu} \! = & 2\lambda_\nu\lambda_\mu
		\int_{-\infty}^{\infty} \!\!\!\!\! \d t  
		\int_{-\infty}^{\infty} \!\!\!\!\! \d t' \!
		\int  \d^d \bm{x}  
		\int  \d^d \bm{x}' \,
		\label{quadL}\\ & \times\notag
		L^*_{\gamma} (t ,\bm{x} ) \,
		 L_{\nu} (t'\!,\bm{x}') \, 
		\left(W_{{\Omega_1}}^{:\phi^2:} (t,\bm{x},t'\!,\bm{x}')\right)^2 
\end{align}
where  $\mathcal{M}_{\gamma\nu}$ has been shown to exhibit persistent divergences for pairs of detectors with the same energy gap \cite{sachs:2017aa}, while $\mathcal{L}_{\gamma\gamma}$ has been shown to be free of persistent divergences \cite{hummer:2016aa}.

\subsection{Quadratic coupling to single complex scalar field\label{sec:complex}}

In this section consider a UDW-like model coupled to a complex scalar field,  previously studied   \cite{hummer:2016aa} in comparison to the fermionic model.  One question that later emerged  \cite{sachsThesis} was whether or not the same persistent divergences found in the quadratic model \cite{sachs:2017aa} also appeared in this model. We will study that relationship further in Section~\ref{sec:discussion}.

The interaction Hamiltonian for a charged scalar field $\Phi$ is
\begin{align}
    \op H(t)=\sum_{\gamma\in\{A,B\}}\int \d^d\vec{x}\op D_\gamma(\vec{x},t)
    :\op \Phi^\dagger(\vec x, t) \op \Phi(\vec x, t):
\label{complexfieldham}
\end{align}
where we now have  
\begin{align}
   f^\gamma_1(\op\Phi_1(\vec x,t))=:\op\Phi_1(\vec x,t)\op\Phi^\dagger_1(\vec x,t):
\end{align}
where the colon notation still indicates normal ordering.

In Appendix~\ref{ap:complex}, we find an expression for $W_{\Omega_1}^{:\op\Phi\op\Phi^\dagger:}(\vec x, t,\vec x', t')$ indicating that the complex correlation function is related to the Wightman function of a real scalar field via
\begin{align}  
        W^{:\Phi\Phi^\dagger:}(t,\bm{x},t' ,\bm{x}')=&
        \left(\bra{0} 
            \op\phi(\bm{x},t)  
            \op\phi (\bm{x}',t')
        \ket{0}\right)^2.\label{complexWightmanthingy2}
\end{align}

Given this relationship, the two-detector density matrix after this interaction is characterized once again by the expressions \cite{sachsThesis}
\begin{align} 
	\mathcal{M}^{:\Phi\Phi^\dagger:}_{\gamma\nu} \! =	&-\lambda_{\aa}\lambda_{\bb}
		\int_{-\infty}^{\infty} \!\!\!\!\! \d t  
		\int_{-\infty}^{t} \!\!\!\!\! \d t' \!
		\int \d^d \bm{x}  
		\int  \d^d \bm{x}' \, 
		\notag\\ & \times\label{complexM}
		L_{\gamma} (t,\bm{x})
        L_{\nu} (t'\!,\bm{x}') \,
		\left(W_{{\Omega_1}}^{ \phi} (t,\bm{x},t'\!,\bm{x}')\right)^2\\[3mm]
	\mathcal{L}^{:\Phi\Phi^\dagger:}_{\gamma\nu} \! = & \lambda_\nu\lambda_\mu
		\int_{-\infty}^{\infty} \!\!\!\!\! \d t  
		\int_{-\infty}^{\infty} \!\!\!\!\! \d t' \!
		\int  \d^d \bm{x}  
		\int  \d^d \bm{x}' \,
		\label{complexL}\\ & \times\notag
		L^*_{\gamma} (t ,\bm{x} ) \,
		 L_{\nu} (t'\!,\bm{x}') \, 
		\left(W_{{\Omega_1}}^{ \phi} (t,\bm{x},t'\!,\bm{x}')\right)^2.
\end{align}  
We can immediately see that the complex model has a close relationship with the quadratically coupled model. We consider this further in Section~\ref{sec:discussion}.

\subsection{Multi-linear coupling to real scalar fields\label{sec:bilinear}}
In this section, we will derive an expression for the state of two arbitrary detectors coupled linearly to arbitrarily many real scalar fields. Unlike the previous two models, this UDW-like model has not been studied before. This is in part because the non-linearly coupled models were introduced in the context of building an UDW-like fermionic model and the multi-linear model is not well-suited for comparison to the fermionic model. However, the divergent behaviour found in \cite{sachs:2017aa} motivates studying models related to the divergent one, partly to determine how pervasive the divergence is to quadratic models and partly as an exploration into potential regularization schemes.

The Hamiltonian for the UDW-like multi-linear model is the following
\begin{align}
    \op H(t)=\sum_{\gamma\in\{A,B\}}\int \d^d\vec{x}\op D_\gamma(\vec{x},t)
	\, \prod_{i=1}^{n}
    \op \phi_i(\vec x, t),
\label{bilinfieldham}
\end{align}
which we have oibtained from the more general \Eq{multifieldham} by letting  $f^\gamma_i(\op\phi_i(\vec x,t))=  \op \phi_i(\vec x, t)$ for $i=1,\ldots, n$. We find the following expressions 
\begin{align} 
	\mathcal{M}^{\phi_1\ldots\phi_n}_{\gamma\nu} \! =	&-\lambda_{\aa}\lambda_{\bb}
		\int_{-\infty}^{\infty} \!\!\!\!\! \d t  
		\int_{-\infty}^{t} \!\!\!\!\! \d t' \!
		\int \d^d \bm{x}  
		\int  \d^d \bm{x}' \, 
		\notag\\ & \times\label{bilinearM}
		L_{\gamma} (t,\bm{x})
        L_{\nu} (t'\!,\bm{x}') \,
        \prod_{i=1}^n
		W_{\ket{\Omega_i}}^{ \phi_i} (t,\bm{x},t'\!,\bm{x}')\\[3mm]
	\mathcal{L}^{\phi_1\ldots\phi_n}_{\gamma\nu} \! = & \lambda_\nu\lambda_\mu
		\int_{-\infty}^{\infty} \!\!\!\!\! \d t  
		\int_{-\infty}^{\infty} \!\!\!\!\! \d t' \!
		\int  \d^d \bm{x}  
		\int  \d^d \bm{x}' \,
		\label{bilinearL}\\ & \times\notag
		L^*_{\gamma} (t ,\bm{x} ) \,
		 L_{\nu} (t'\!,\bm{x}') \, 
        \prod_{i=1}^n
		W_{\ket{\Omega_i}}^{ \phi_i} (t,\bm{x},t'\!,\bm{x}')
\end{align}
that characterize the two-detector density matrix after a multi-linear interaction with the scalar fields.
We see in the above $n$th powers of the Wightman function. This is particularly interesting in the case of $n=2$, as the resultant expressions are exactly those found for the UDW-like model coupled to a complex field. This similarity and other insights will be discussed in Section~\ref{sec:discussion}.


\section{Discussion\label{sec:discussion}}

In this section, we will analyze and compare the results derived from the four models in the previous section. Relying on the previous study \cite{sachs:2017aa}, we will draw conclusions about the divergent behaviour (if any) of the   models' predictions,  considering in particular   \emph{persistent} divergences in the correlation terms. Then, we will discuss the nature of the divergences as they relate to distribution theory and operator algebras, which will also have bearing on possible regularization schemes. This sets the stage for a later discussion (Section \ref{sec:futurework})   of these results in the context of possible routes to regularization. 

The first observation we make is that the forms of $\mathcal{M}^{\phi_1\phi_2}$ and $\mathcal{M}^{:\Phi\Phi^\dagger:}$ are precisely the same. Furthermore these two expressions differ from $\mathcal{M}^{:\phi^2:}$ by only a constant factor. Thus in any case where the real quadratic model presents persistent divergences, so too will the complex  and bilinear models. Concretely, these models will exhibit the same persistent divergences as the quadratically coupled detector model studied in \cite{sachs:2017aa}. This will lead to divergent negativity unless regularization schemes are employed or a satisfactory renormalzation technique is found. 
Furthermore, we can conclude that the regularization methods or renormalization techniques used to tame the divergences for the quadratic model should work equally well for the complex scalar field, or for a double coupling to two different real scalar fields. 

Our final observation is about the nature of the higher-order multi-linear interactions. While the linear interaction is free of persistent divergences, and the bi-linear case has now been shown to exhibit persistent divergences, in order to draw conclusions about the divergence structure of the multi-linear UDW-like models we cannot immediately appeal to the behaviour of higher powers of the Wightman function. For that, we need to discuss the divergence in terms of distribution theory.

We cannot yet answer here the question of whether the multi-linear models with more than two fields do exhibit persistent divergences, but it is productive to point out that, even for the quadratic model, it might perhaps be surprising that any of the density matrix elements of the quadratic models are convergent at all. This is because the role that the Wightman function plays for a linear detector in the detector's matrix elements is played by \emph{products} of Wightman functions in the quadratic and bilinear cases. However, the Wightman function is a distribution, and the product of two distributions that are well-defined in a reasonable test space is not guaranteed to be a well-defined distribution on the same space.

Finally, another way of looking at this issue of divergences is via considering algebras of operators. Smeared field operators are well-defined objects \cite{wald}. However, a smeared \emph{product} of field operators (as we see in the quadratic models studied in this paper)  is not, a-priori, a well-defined object in the algebra of field operators. One method to give the quadratic Hamiltonian a well-defined interpretations is as a limit of well-defined non-local Hamiltonians, as in
\begin{align}
    \int \d y\int \d x F(\vec x) \op\phi(\vec x)   G_\delta(\vec y-\vec x) \op\phi(\vec y),
\end{align}
where $G_\delta(\vec x)$ is a nascent delta distribution.

A more intriguing  alternative regularization scheme might exploit recent work
\cite{ng:2018aa} demonstrating that  for general spacetimes and spacelike-separated pairs of detectors following arbitrary timelike trajectories, it is possible to de-nest the nested integrals in $\mathcal{M}$ and write them as functions of slightly-altered $\mathcal{L}_{\gamma\nu}$.   Since all functions $\mathcal{L}_{\gamma\nu}$ for pairs of quadratically coupled detectors have been found to be free of persistent divergences, it may be possible regularize the quadratic models through analyzing this de-nesting process.

A final, perhaps more elegant, regularization scheme (still different from a UV cutoff) is applying \emph{multiple smearings}, one for each instance of the field operator in the interaction Hamiltonian. Under this regularization scheme, the smeared field operators would be rigorously defined in terms of operator algebas. More concretely, every field operator appearing in the Hamiltonian is smeared under an integral separately. For example, instead of the multi-linear Hamiltonian defined in \Eq{bilinfieldham},
\begin{align}
    \op H(t)=\sum_{\gamma\in\{A,B\}}\int \d^d\vec{x}\op D_\gamma(\vec{x},t)
	\, \prod_{i=1}^{n}
    \op \phi_i(\vec x, t),
\end{align}
one could use the following multiply-smeared detector
\begin{align}
    \op H(t)=
    \!\!\sum_{\gamma\in\{A,B\}}\!\!
    \op D_\gamma(t)
	\, \prod_{i=1}^{n}
    \int \d^d\vec{x}_i\,
    F_{i,\gamma}(\vec x_i)\,
    \op \phi_i(\vec x_i, t),
\label{multismearedham}
\end{align}
where $F_{i,\gamma}$ is the smearing associated with the $i$th field and the detector $\gamma$.

\section{Conclusion}
\label{sec:futurework}

We have derived expressions for the time evolved state of a pair of Unruh-DeWitt detectors coupling to various scalar fields in different configurations.  We have shown that the phenomenlology associated with the vacuum response of pairs of detectors at leading order (in, for example, entanglement harvesting) is essentially insensitive (modulo a constant factor) to the charged nature of the field, as well as the presence of two different types of scalar fields in the coupling. 

  Regarding persistent divergences, by comparing a pair of Unruh-DeWitt detectors with  multilinear coupling to $n$ real scalar fields, we find for $n=2$ that the multi-linear model exhibits the same divergences as the quadratically coupled model studied previously \cite{hummer:2016aa,sachs:2017aa}.  The divergence structure is the same for a  pair of Unruh-DeWitt detectors coupled quadratically to a complex field.  

Though we leave for future investigation the question of how to treat these persistent divergences,  we  close with  a  few possible thoughts on this: Given the discussion in Section \ref{sec:discussion}, a regularization scheme beyond the soft UV cutoff employed in \cite{sachs:2017aa} may perhaps be found through analyzing the distributional nature of the Wightman function, applying point-splitting techniques,   regularizing through field observable-localization, or exploiting recent results in \cite{ng:2018aa} whose analysis suggests that perhaps these divergencies can be regularized through frequency detuning. These are very promising avenues that will be explored in future investigations.

\begin{acknowledgments}
A.M.S. would like to thank Emma M McKay and Jose ``Pipo" de Ramon Rivera for helpful discussion. This work was supported in part by the Natural Sciences and Engineering Research Council of Canada   through the Discovery program. E.M-M. also acknowledges the support of his Ontario Early Researcher Award.
\end{acknowledgments}

\appendix
\section{Two-point correlator for the complex field\label{ap:complex}}

In this appendix, we find an expression for $W_{\Omega_1}^{:\op\Phi\op\Phi^\dagger:}(\vec x, t,\vec x', t')$. Concretely, we will relate the complex correlation function to the usual Wightman function as shown in \Eq{complexWightmanthingy2}.

We begin with the relationship between an operator $\op A$ and its normal ordered version is given by
\begin{align}  
	  :   \op A   : \,= \op A - \bra{0}\! \op A \ket{0}{\edu \openone} 
\end{align}
Using this identity, $W^{:\op\Phi\op\Phi^\dagger:}$ can be rewritten as
\begin{align}  
    W^{:\op\Phi\op\Phi^\dagger:}&(t,\bm{x},t' ,\bm{x}') =\label{thingBeforePhiCommiescomplex}
    \\&
	    \bra{0} 
            \op\Phi (\bm{x},t) \op\Phi ^\dagger(\bm{x},t)
            \op\Phi (\bm{x}',t') \op\Phi ^\dagger(\bm{x}',t') 
        \ket{0}\notag\\
        &- 
        \bra{0} 
            \op\Phi (\bm{x},t) \op\Phi ^\dagger(\bm{x},t)
        \ket{0} 
        \bra{0} 
            \op\Phi (\bm{x}',t') \op\Phi ^\dagger(\bm{x}',t') 
        \ket{0}.\notag
\end{align}
The first term of $W^{:\op\Phi\op\Phi^\dagger:}(t,\bm{x},t' ,\bm{x}')$ can be simplified. To do so, we will write the field operator as $\op\Phi=\op\Phi^++\op\Phi^-$, where $\op\Phi^+$ and $\op\Phi^-$ are defined as
\begin{align}  
	\op\Phi^+ (\bm{x},t) &=
	    \int  \frac{\d^d \bm{k} \, 
		e^{- \epsilon |\bm{k}|/2}}{\sqrt{2(2\pi)^n|\bm{k}|}}\,
		    \hat{a}^\dagger_k \,
		    e^{\ii (|\bm{k}|t-\bm{k}\cdot\bm{x})}     ,\\
    \op\Phi^- (\bm{x},t) &=
		\int  \frac{\d^d \bm{k} \, 
		e^{- \epsilon |\bm{k}|/2}}{\sqrt{2(2\pi)^n|\bm{k}|}} \,
		    \hat{b}^{\phantom{\dagger}}_k \,
		    e^{-\ii (|\bm{k}|t-\bm{k}\cdot\bm{x})},
\end{align}
where the operators $\op\Phi^+$ and $\op\Phi^-$ satisfy the commutation relation
\begin{align}  
    \big[ \op\Phi^- (\bm{x}_\mu,t_\mu) ,  \op\Phi^+ (\bm{x}_\nu,t_\nu) \big]=
    0
    ,\label{comrelation1}
\end{align}
and 
\begin{align}  
    &\big[ \op\Phi^- (\bm{x}_\mu,t_\mu) ,  \op\Phi^{-\dagger} (\bm{x}_\nu,t_\nu) \big]=
    \mathcal{C}_{\mu\nu}\openone
    \label{comrelation2complex}
    ,\\
    &\big[ \op\Phi^+ (\bm{x}_\mu,t_\mu) ,  \op\Phi^{+\dagger} (\bm{x}_\nu,t_\nu) \big]=
    \mathcal{C}_{\nu\mu}\openone,
\end{align}
where $\mathcal{C}_{\mu\nu}\in \mathbb{C}$ is precisely
\begin{align}  
    \mathcal{C}_{\mu\nu}=&
		\int  \frac{\d^d \bm{k} \, e^{- \epsilon |\bm{k}|/2}}{2(2\pi)^n|\bm{k}|}
		e^{\ii (|\bm{k}|(t_\nu-t_\mu)-\bm{k}\cdot(\bm{x}_\nu-\bm{x}_\mu))}\notag
\end{align}

Furthermore,   to simplify notation  we define $\op\Phi_\nu$ such that 
\begin{align}  
    \bra{0} &
            \op\Phi  (\bm{x},t) 
            \op\Phi ^\dagger(\bm{x},t)
            \op\Phi  (\bm{x}',t') 
            \op\Phi ^\dagger(\bm{x}',t') 
        \ket{0}
        =\notag\\&
        \bra{0} 
            \op\Phi _1^{\phantom{\dagger}}
            \op\Phi _2^\dagger
            \op\Phi _3^{\phantom{\dagger}}
            \op\Phi _4^\dagger
             \ket{0},\label{tempdefines}
\end{align}
we can use $\op\Phi^+$ and $\op\Phi^-$ (and their adjoints) to rewrite the first term in Eq.~\eqref{thingBeforePhiCommiescomplex}  as
\begin{align}  
        \bra{0}& 
            \op\Phi_1^{\phantom{\dagger}}
            \op\Phi_2^\dagger      
            \op\Phi_3^{\phantom{\dagger}}
            \op\Phi_4^\dagger      
        \ket{0}=
        \bra{0}  
            \op\Phi^-_1   
            \op\Phi_2^{-\dagger}
            \op\Phi^-_3   
            \op\Phi_4^{-\dagger}   
        \ket{0}\notag\\&+
        \bra{0}  
            \op\Phi^-_1    
            \op\Phi^{+\dagger}_2   
            \op\Phi^-_3   
            \op\Phi^{+\dagger}_4   \ket{0}\label{peeeeepobastardcomplex}.
\end{align}
Note that here the expression differs from the real scalar field case (see equation (A7) of \cite{sachs:2017aa}), although we still use that
\begin{equation}
{\op\Phi}_\mu^{+\dagger}           \ket{0}
=
\bra{0}     {\op\Phi}_\nu^+
=
{\op\Phi}_\mu^-   \ket{0}
=
\bra{0}     {\op\Phi}_\nu^{-\dagger}
=0,
\end{equation}
and that only summands with as many $\op\Phi^-$ as $\op\Phi^{-\dagger}$ and $\op\Phi^+$ as $\op\Phi^{+\dagger}$ give a non-vanishing vacuum expectation

By commuting operators using Eq.~\eqref{comrelation1}, we can write the the following 
\begin{align}  
        \bra{0}  
            \op\Phi^-_1   
            \op\Phi_2^{-\dagger}   
            \op\Phi^-_3   
            \op\Phi_4^{-\dagger}   
            \ket{0}
            &=
        \mathcal{C}_{23}\mathcal{C}_{14}   
    \notag\\
    \bra{0}  
        \op\Phi^-_1    
        \op\Phi^+_2   
        \op\Phi^-_3   
        \op\Phi^+_4   
    \ket{0}
    &=\mathcal{C}_{12}\mathcal{C}_{34} .
\end{align}
Thus Eq.~\eqref{peeeeepobastardcomplex} can be written as
\begin{align}  
        \bra{0}& 
        \op\Phi_1  ^{\phantom{\dagger}}
        \op\Phi_2 ^{\phantom{\dagger}}
        \op\Phi_3 ^{\phantom{\dagger}}
        \op\Phi_4 ^{\phantom{\dagger}}
        \ket{0}=
        \mathcal{C}_{23}\mathcal{C}_{14} +
        \mathcal{C}_{12}\mathcal{C}_{34}.\label{OMGtheDeltascomplex}
\end{align}
We can rewrite the $\mathcal{C}_{\mu\nu}$ coefficients as 
\begin{align}  
    \mathcal{C}_{\mu\nu}=
    \bra{0}
        \big[ \op\Phi_\mu^-,\op\Phi_\nu^{-\dagger} \big]
    \ket{0}
    =&
    \bra{0} 
        \op\Phi_\mu^-\op\Phi_\nu^{-\dagger}
    \ket{0}
    =
    \bra{0}
        \op\Phi_\mu\op\Phi_\nu^\dagger
    \ket{0},
\end{align}
which allows us to write the following relation
\begin{align}  
        \bra{0} \op\Phi_1^{\phantom{\dagger}}
        \op\Phi_2^{\phantom{\dagger}}
        \op\Phi_3^{\phantom{\dagger}}
        \op\Phi_4^{\phantom{\dagger}}
        \ket{0}=&
        \bra{0} 
        \op\Phi_1^{\phantom{\dagger}}
        \op\Phi_2^\dagger
        \ket{0}
        \bra{0} 
        \op\Phi_3^{\phantom{\dagger}}
        \op\Phi_4^\dagger 
        \ket{0}\notag\\&
        +\bra{0} 
        \op\Phi_1^{\phantom{\dagger}}
        \op\Phi_3^\dagger
        \ket{0}
        \bra{0} 
        \op\Phi_2^{\phantom{\dagger}}
        \op\Phi_4^\dagger 
        \ket{0}.
\end{align}
Using our definition of $\op\Phi_\nu$ in Eq.~\eqref{tempdefines}, this becomes
\begin{align}  
        \bra{0} &\op\Phi(\bm{x},t)  \op\Phi^\dagger(\bm{x},t) \op\Phi(t' ,\bm{x}') \op\Phi^\dagger(t' ,\bm{x}') \ket{0}=
        \notag\\&
        \bra{0}
            \op\Phi(\bm{x},t)  
            \op\Phi^\dagger(\bm{x},t)\ket{0}
        \bra{0}
            \op\Phi(t' ,\bm{x}') 
            \op\Phi^\dagger(t' ,\bm{x}') \ket{0}
        \notag\\&
        +\bra{0}
            \op\Phi(\bm{x},t)  
            \op\Phi^\dagger(t' ,\bm{x}')\ket{0}
        \bra{0}
            \op\Phi(\bm{x},t) 
            \op\Phi^\dagger(t' ,\bm{x}') \ket{0}.
\end{align}

Thus the complex correlation function, originally expressed in Eq.~\eqref{thingBeforePhiCommiescomplex}, is
\begin{align}  
        W^{:\Phi\Phi^\dagger:}(t,\bm{x},t' ,\bm{x}')=&
        \left(\bra{0} 
            \op\Phi(\bm{x},t)  
            \op\Phi^\dagger (\bm{x}',t')
        \ket{0}\right)^2.\label{complexWightmanthingy1}
\end{align}

\bibliography{bib}

\end{document}